\documentclass[12pt]{article}

\makeatletter
 
\def\section{\@startsection {section}{1}{\z@}{+3.0ex plus +1ex minus
  +.2ex}{2.3ex plus .2ex}{\normalsize\bf}}
\def\subsection{\@startsection{subsection}{2}{\z@}{+2.5ex plus +1ex
minus +.2ex}{1.5ex plus .2ex}{\normalsize\bf}}
\def\subsubsection{\@startsection{subsubsection}{3}{\z@}{+3.25ex plus
 +1ex minus +.2ex}{1.5ex plus .2ex}{\normalsize\bf}}
 
\expandafter\ifx\csname mathrm\endcsname\relax\def\mathrm#1{{\rm #1}}\fi
 
\makeatletter

\newcount\@tempcntc
\def\@citex[#1]#2{\if@filesw\immediate\write\@auxout{\string\citation{#2}}\fi
  \@tempcnta\z@\@tempcntb\m@ne\def\@citea{}\@cite{\@for\@citeb:=#2\do
    {\@ifundefined
       {b@\@citeb}{\@citeo\@tempcntb\m@ne\@citea
        \def\@citea{,\penalty\@m\ }{\bf ?}\@warning
       {Citation `\@citeb' on page \thepage \space undefined}}%
    {\setbox\z@\hbox{\global\@tempcntc0\csname
b@\@citeb\endcsname\relax}%
     \ifnum\@tempcntc=\z@ \@citeo\@tempcntb\m@ne
       \@citea\def\@citea{,\penalty\@m}
       \hbox{\csname b@\@citeb\endcsname}%
     \else
      \advance\@tempcntb\@ne
      \ifnum\@tempcntb=\@tempcntc
      \else\advance\@tempcntb\m@ne\@citeo
      \@tempcnta\@tempcntc\@tempcntb\@tempcntc\fi\fi}}\@citeo}{#1}}

\def\@citeo{\ifnum\@tempcnta>\@tempcntb\else\@citea
  \def\@citea{,\penalty\@m}%
  \ifnum\@tempcnta=\@tempcntb\the\@tempcnta\else
   {\advance\@tempcnta\@ne\ifnum\@tempcnta=\@tempcntb \else
\def\@citea{--}\fi
    \advance\@tempcnta\m@ne\the\@tempcnta\@citea\the\@tempcntb}\fi\fi}

\def\asymp#1%
{\mathrel{\raisebox{-.4em}{$\widetilde{\scriptstyle #1}$}}}

\def\Nequal#1%
{\mathrel{\raisebox{-.5em}{$\stackrel{=}{\scriptstyle\rm#1}$}}}

\def\beq#1\eeq{\begin{equation}#1\end{equation}}
\def\beqar{\begin{eqnarray}}
\def\eeqar{\end{eqnarray}}
\def\barr#1{\begin{array}{#1}}
\def\earr{\end{array}}
\def\bfi{\begin{figure}}
\def\efi{\end{figure}}
\def\btab{\begin{table}}
\def\etab{\end{table}}
\def\bce{\begin{center}}
\def\ece{\end{center}}
\def\nn{\nonumber}

\def\veps{\varepsilon}

\def\refeq#1{\mbox{(\ref{#1})}}

\def\refse#1{\mbox{Section~\ref{#1}}}

\def\citere#1{\mbox{Ref.~\cite{#1}}}
\def\citeres#1{\mbox{Refs.~\cite{#1}}}

\newcommand{\ri}{{\mathrm{i}}}

\newcommand{\rd}{{\mathrm{d}}}

\newcommand{\rL}{{\mathrm{L}}}

\newcommand{\rT}{{\mathrm{T}}}

\def\mathswitchr#1{\relax\ifmmode{\mathrm{#1}}\else$\mathrm{#1}$\fi}

\newcommand{\PW}{\mathswitchr W}

\newcommand{\PS}{\mathswitchr S}
\newcommand{\Pe}{\mathswitchr e}

\newcommand{\Pem}{\mathswitchr {e^-}}
\newcommand{\PWp}{\mathswitchr {W^+}}

\def\mathswitch#1{\relax\ifmmode#1\else$#1$\fi}

\newcommand{\MW}{\mathswitch {M_\PW}}

\newcommand{\Me}{\mathswitch {m_\Pe}}

\def\Re{\mathop{\mathrm{Re}}\nolimits}

\newcommand{\z}{\setbox0\hbox{+}\hbox to \wd0{\hss0\hss}}
\def\limfunc#1{\mathop{\rm #1}}

\def\Re{\limfunc{Re}}

\def\slash#1{{\setbox0=\hbox{$#1$}
  \rlap{\ifdim\wd0>.7em\kern.22\wd0\else\kern.1\wd0\fi /}#1}}

\def\braket#1#2{\left\langle #1\vphantom{#2}
  \right. \kern-2.5pt\left| #2\vphantom{#1}\right\rangle }

\def\What{{\hat W}}

\def\Zhat{{\hat Z}}
\def\Ahat{{\hat A}}

\def\Hhat{\hat H}

\def\phihat{\hat \phi}

\def\chihat{\hat \chi}
\def\xihat{\hat \xi}

\def\rT{{\mathrm{T}}}

\def\rL{{\mathrm{L}}}

\begin{document}

\thispagestyle{empty}
\def\thefootnote{\fnsymbol{footnote}}
\setcounter{footnote}{1}
\null
\strut\hfill CERN-TH/97-78 \\
\strut\hfill hep-ph/9704368
\vskip 0cm
\vfill
\begin{center}
{\Large \bf 
\boldmath{Thirring's low-energy theorem and its generalizations\\[.5em]
in the electroweak Standard Model 
}%
\par} \vskip 2.5em
{\large
{\sc Stefan Dittmaier}\\[1ex]
{\normalsize \it Theory Division, CERN\\
CH-1211 Geneva 23, Switzerland}
}
\par \vskip 1em
\end{center} \par
\vskip 3cm 
{\bf Abstract:} \par
The radiative corrections to Compton scattering vanish in the low-energy
limit in all orders of perturbation theory.
This theorem, which is well-known for Abelian gauge theories, is
proved in the electroweak Standard Model. Moreover, analogous theorems
are derived for photon scattering off other charged particles, in
particular W~bosons. Since the theorems follow from
gauge invariance and on-shell renormalization, their derivation is most
conveniently performed in the framework of the background-field method. 
\par
\vfill
\noindent 
CERN-TH/97-78 \\
April 1997 \par
\vskip 1cm 
\null
\setcounter{page}{0}
\clearpage
\def\thefootnote{\arabic{footnote}}
\setcounter{footnote}{0}

\section{Introduction}

In all models for particle interactions that contain QED, the electric
charge $-e$ of the electron is usually fixed in the Thomson limit,
i.e.\ by low-energy Compton scattering, $\Pem\gamma\to\Pem\gamma$.
This definition ensures that $e$ indeed represents the electric charge
that describes the interaction between electrons and the electromagnetic
field in classical electrodynamics.
However, in the commonly adopted on-shell renormalization scheme,
$e$ is formally defined as the coupling of a photon
with momentum zero to an on-shell electron. In QED the equivalence of
the two definitions of $e$ was first proved by Thirring \cite{th50},
who showed that all radiative corrections to Compton scattering vanish
in the low-energy limit within the on-shell scheme. Subsequently, this
theorem was rederived and modified for more general U(1) gauge theories
by several authors \cite{kr54}. The validity of the theorem is commonly
assumed also for theories like the electroweak Standard Model (SM),
where QED is embedded in a non-Abelian gauge symmetry, 
although---to the best of our knowledge---no field-theoretical proof has
been given so far. The purpose of this paper is to close this gap for
the SM and to generalize the theorem to elastic low-energy 
scattering between photons and charged particles other than electrons, 
for instance to $\PWp\gamma\to\PWp\gamma$.

Thirring's theorem is a consequence of on-shell renormalization and gauge 
invariance, which implies relations between Green functions. In QED, these
relations are the well-known Ward identities. In non-Abelian gauge
theories the BRS symmetry, which replaces the original gauge symmetry
after the usual Faddeev--Popov quantization, leads to much more
complicated relations. These are known as Slavnov--Taylor identities, which
in general involve explicit contributions of Faddeev--Popov ghost fields. 
In particular, the BRS symmetry of the SM does not lead to those Ward 
identities 
that imply Thirring's theorem in QED. This is due to the fact that the
electromagnetic current is not strictly conserved in the non-Abelian
case. More precisely, the divergence of this current acts as the null 
operator only on physical states, because it is the BRS
transform of an appropriate operator. 
Therefore, the derivations of the low-energy theorems presented in
\citere{kr54}, which rely on strict electromagnetic current
conservation, are not valid in theories such as the SM. 
In such cases a proof has to be based on the underlying BRS symmetry 
in the conventional Faddeev--Popov approach.
However, if non-Abelian gauge theories are quantized in the
framework of the background-field method (BFM) 
\cite{BFMref,Ab81,Ab83,bgflett,bgflong,li95,bgfet},
the vertex and Green functions obey simple QED-like Ward identities.
Since this feature greatly simplifies the investigation of the
consequences of gauge invariance, we adopt this formalism in the
following. 

The outline of the article is as follows:
After some preliminary remarks about our calculational framework in
\refse{se:prel},
we inspect the low-energy limit of Compton scattering within the SM
in \refse{se:eaea}. In \refse{se:other} we show how the treatment of
the electron carries over to other charged particles, where particular
attention is paid to the W~boson. A summary is presented in
\refse{se:sum}.

\section{Preliminary remarks}
\label{se:prel}

As the BFM is not yet as well-known as the conventional
quantization procedure, we first make some brief remarks on the
basis and the starting point of our analysis. 
The BFM represents an alternative formalism for the quantization of 
non-Abelian gauge theories, which was invented in order to retain the 
gauge invariance of the effective action. 
The method was developed in different ways in
\citeres{BFMref,Ab81,Ab83} for pure, massless Yang--Mills theories and
more recently applied to the SM in \citeres{bgflett,bgflong,li95,bgfet},
where Abbott's formulation \cite{Ab81} was used as guideline.

The first step in the BFM approach is the construction of an effective 
action that is invariant under gauge transformations.
To this end, each field in the classical Lagrangian is split
into a background part and a quantum part, where only the latter is
quantized, and the former appears as auxiliary field.
The Faddeev--Popov quantization of the
quantum fields requires the fixing of their gauge.
The corresponding gauge-fixing conditions are chosen such that the 
complete path integral 
remains invariant under gauge transformations of the background fields. 
The background fields are identified with the independent variables of 
the BFM effective action, which is thus gauge-invariant.
Diagrammatically this means that quantum fields appear on internal
lines of
one-particle-irreducible loop diagrams, whereas background fields
represent the external legs. The gauge invariance of the BFM effective
action implies that the vertex functions, which follow from this
effective action by taking functional derivatives, obey the simple Ward
identities that are related to the classical Lagrangian. 
In \citeres{bgflong,li95} it was shown that the BFM Ward identities are
compatible with the on-shell renormalization of the SM when
the field renormalization is chosen appropriately. In the following all
vertex functions $\Gamma^{\cdots}$ and self-energies
$\Sigma^{\cdots}$ are assumed to be renormalized according to
the scheme of \citere{bgflong}.

One-particle-reducible connected Green functions and $S$-matrix elements 
are constructed by forming trees with the (one-particle-irreducible) 
vertex functions joined by propagators for the background fields \cite{Ab83}.
The background-field propagators are defined by introducing a
gauge-fixing term for the background gauge fields, 
which is, however, not related to the one that fixes the gauge
of the quantum fields.
The associated additional gauge parameters $\hat\xi_{\cdots}$
enter only via tree-level quantities, but not via the higher-order
contributions to the vertex functions.
Therefore, the simple Ward identities for the vertex functions
$\Gamma^{\cdots}$ translate into simple relations between
connected Green functions $G^{\cdots}$. For a 't~Hooft
gauge-fixing term for the background fields in the SM, these identities were
explicitly derived in \citere{bgfet}.

Our notations and conventions for vertex and Green functions exactly follow
the ones of \citeres{bgflett,bgflong,bgfet}, i.e.\ for instance all
labelled fields and momenta are incoming. Background fields are marked by 
carets, except for fermion fields, where background
and quantum fields need not be distinguished. 

\section{Low-energy Compton scattering---Thomson limit} 
\label{se:eaea}

In order to prove the low-energy theorem for 
Compton scattering, $\Pem\gamma\to\Pem\gamma$, we start by deriving the
Ward identity for the corresponding connected Green function
$G^{\Ahat\Ahat\bar\Pe\Pe}_{\mu\nu}$, where both photon legs are
contracted with their momenta. This identity is easily obtained from the
first relation of Eq.~(11) in \citere{bgfet}, which is the identity for
the generating functional expressing electromagnetic gauge invariance.
Taking derivatives with respect to the sources of the fields 
$\bar\Pe$, $\Pe$ and $\Ahat$, $\bar\Pe$, $\Pe$, respectively, in
momentum space one gets
\beqar
\frac{\ri k^2}{\xihat_A}\,k^\alpha G^{\Ahat\bar\Pe\Pe}_{\alpha}(k,\bar p,p) 
&=& e G^{\bar\Pe\Pe}(\bar p,-\bar p) - e G^{\bar\Pe\Pe}(-p,p),
\label{eq:kGAee}
\\
\frac{\ri k_1^2}{\xihat_A}\,k_1^\mu 
G^{\Ahat\Ahat\bar\Pe\Pe}_{\mu\nu}(k_1,k_2,\bar p,p) 
&=& e G^{\Ahat\bar\Pe\Pe}_{\nu}(k_2,\bar p,-\bar p-k_2) 
- e G^{\Ahat\bar\Pe\Pe}_{\nu}(k_2,-p-k_2,p).
\label{eq:kGAAee1}
\eeqar
Upon contracting \refeq{eq:kGAAee1} with $k_2^\nu$ and using
\refeq{eq:kGAee}, one obtains the desired identity,
\beqar
\lefteqn{
\frac{\ri k_1^2}{\xihat_A}\, \frac{\ri k_2^2}{\xihat_A}\, k_1^\mu k_2^\nu 
G^{\Ahat\Ahat\bar\Pe\Pe}_{\mu\nu}(k_1,k_2,\bar p,p) 
} && \nn\\ && =
e^2 G^{\bar\Pe\Pe}(\bar p,-\bar p) - 
e^2 G^{\bar\Pe\Pe}(\bar p+k_1,p+k_2) - 
e^2 G^{\bar\Pe\Pe}(\bar p+k_2,p+k_1) + 
e^2 G^{\bar\Pe\Pe}(-p,p).
\label{eq:kGAAee2}
\eeqar

The next step consists in amputating the external photon propagators in
$G^{\Ahat\Ahat\bar\Pe\Pe}_{\mu\nu}$. Denoting amputated external fields
by a lowered field index, we have
\beq
G^{\Ahat\cdots}_\alpha (k,\ldots) = 
\sum_X G^{\Ahat X}_\alpha (k,-k) G^{\cdots}_X(k,\ldots),
\label{eq:amp}
\eeq
where $X$ stands for all (background) fields that can mix with the 
(background) photon field $\Ahat$. In the SM the sum over $X$ 
extends over $\Ahat$, $\Zhat$, $\Hhat$, and $\chihat$. Substituting
$G^{\Ahat\Ahat\bar\Pe\Pe}_{\mu\nu}$ in the l.h.s.\ of \refeq{eq:kGAAee2} 
according to \refeq{eq:amp}, and 
using the Ward identities for the two-point functions 
$G^{\Ahat X}_\alpha$ \cite{bgfet},
\beq
k^\alpha G^{\Ahat\Ahat}_{\alpha\beta} (k,-k) = 
-\frac{\ri\xihat_A}{k^2}\,k_\beta, \qquad
k^\alpha G^{\Ahat X}_{\alpha} (k,-k) = 0 \quad \mbox{for} \;
X = \Zhat, \Hhat, \chihat,
\label{eq:GAXwi}
\eeq
we find
\beq
k_{1,\mu} k_{2,\nu} G^{\mu\nu,\bar\Pe\Pe}_{\Ahat\Ahat}(k_1,k_2,\bar p,p) 
= \frac{\ri k_1^2}{\xihat_A}\, \frac{\ri k_2^2}{\xihat_A}\, k_1^\mu k_2^\nu 
G^{\Ahat\Ahat\bar\Pe\Pe}_{\mu\nu}(k_1,k_2,\bar p,p),
\label{eq:kGAAee3}
\eeq
which translates \refeq{eq:kGAAee2} into a Ward identity for 
$G^{\mu\nu,\bar\Pe\Pe}_{\Ahat\Ahat}$.
We note in passing that the counterparts of the BFM Ward identities 
\refeq{eq:kGAAee2}, \refeq{eq:GAXwi}, \refeq{eq:kGAAee3} in the
conventional field-theoretical approach are much more complicated and
involve explicit contributions of Faddeev--Popov ghosts.
In order to get a low-energy limit of
$G^{\mu\nu,\bar\Pe\Pe}_{\Ahat\Ahat}$, we keep the momentum $p$ fixed,
take derivatives from \refeq{eq:kGAAee3} with respect to $k_{1,\mu}$ 
and $k_{2,\nu}$, and take the limit $k_1,k_2 \to 0$. Note that the
momentum $\bar p$ is not independent, but $\bar p=-p-k_1-k_2$.
One obtains
\beq
G^{\mu\nu,\bar\Pe\Pe}_{\Ahat\Ahat}(0,0,-p,p) = 
e^2 \frac{\partial^2}{\partial p_\mu\partial p_\nu}
G^{\bar\Pe\Pe}(-p,p).
\label{eq:GAAee1}
\eeq

For the fully amputated Green function, which is needed for the
$S$-matrix element,
also the electron legs must be amputated by multiplication with the inverse
propagators. Using the relation
between electron propagator $G^{\bar\Pe\Pe}$ and two-point vertex
function $\Gamma^{\bar\Pe\Pe}$,
\beq
G^{\bar\Pe\Pe}(-p,p) \Gamma^{\bar\Pe\Pe}(-p,p) = -1,
\label{eq:Gee}
\eeq
we have
\beqar
G^{\mu\nu}_{\Ahat\Ahat\bar\Pe\Pe}(0,0,-p,p) &=&
e^2 \Gamma^{\bar\Pe\Pe}(-p,p)
\left[ \frac{\partial^2}{\partial p_\mu\partial p_\nu}
G^{\bar\Pe\Pe}(-p,p) \right]
\Gamma^{\bar\Pe\Pe}(-p,p),
\label{eq:GAAee2}
\eeqar
according to \refeq{eq:GAAee1}. The r.h.s.\ of
this equation can explicitly be expressed in terms of the electron
self-energy $\Sigma^{\bar\Pe\Pe}$, which is related to
$\Gamma^{\bar\Pe\Pe}$ by
\beq
\Gamma^{\bar\Pe\Pe}(-p,p) =
\ri\left[ \slash{p}-\Me+\Sigma^{\bar\Pe\Pe}(p) \right] =
\ri\left[ \slash{p}+\slash{p}\omega_\sigma\Sigma^{\bar\Pe\Pe}_\sigma(p^2)
-\Me+\Me\Sigma^{\bar\Pe\Pe}_{\mathrm{S}}(p^2) \right].
\eeq
Here the sum over $\sigma=+/-=\mathrm{R}/\mathrm{L}$ runs over the right-
and left-handed contributions $\Sigma^{\bar\Pe\Pe}_{\mathrm{R}}$ and
$\Sigma^{\bar\Pe\Pe}_{\mathrm{L}}$ to the $\slash{p}$ part of 
$\Sigma^{\bar\Pe\Pe}$, respectively, and $\omega_\pm=\frac{1}{2}(1\pm\gamma_5)$ 
are the chirality projectors. The actual evaluation of \refeq{eq:GAAee2}
is facilitated by shifting the derivatives from $G^{\bar\Pe\Pe}$ to
$\Gamma^{\bar\Pe\Pe}$ with the identity
\beq
\Gamma^{\bar\Pe\Pe}\left[ \frac{\partial^2 G^{\bar\Pe\Pe}}
{\partial p_\mu\partial p_\nu} \right] \Gamma^{\bar\Pe\Pe}
= \left[\frac{\partial\Gamma^{\bar\Pe\Pe}}{\partial p_\mu}\right]
G^{\bar\Pe\Pe}\left[\frac{\partial\Gamma^{\bar\Pe\Pe}}{\partial p_\nu}\right]
+ \left[\frac{\partial\Gamma^{\bar\Pe\Pe}}{\partial p_\nu}\right]
G^{\bar\Pe\Pe}\left[\frac{\partial\Gamma^{\bar\Pe\Pe}}{\partial p_\mu}\right]
+ \frac{\partial^2 \Gamma^{\bar\Pe\Pe}}{\partial p_\mu\partial p_\nu},
\eeq
which can be easily derived by taking derivatives from \refeq{eq:Gee}.
The final result for the amputated Green function 
$G^{\mu\nu}_{\Ahat\Ahat\bar\Pe\Pe}$ with photons of momentum zero reads
\beqar
G^{\mu\nu}_{\Ahat\Ahat\bar\Pe\Pe}(0,0,-p,p) &=&
2e^2 g^{\mu\nu}\left[
\frac{\Gamma^{\bar\Pe\Pe}(-p,p)}
{p^2-\Me^2 f(p^2)}
+\ri\slash{p}\omega_\sigma\Sigma^{\prime\bar\Pe\Pe}_\sigma(p^2)
+\ri\Me\Sigma^{\prime\bar\Pe\Pe}_{\mathrm{S}}(p^2)
\right]
\nn\\[.3em]
&& {} + \;\; \left( \mbox{terms involving $p^\mu$ or $p^\nu$} \right),
\label{eq:GAAee3}
\eeqar
where
\beq
f(p^2) = \frac{\left[1-\Sigma^{\bar\Pe\Pe}_{\mathrm{S}}(p^2)\right]^2}
{\left[1+\Sigma^{\bar\Pe\Pe}_{\mathrm{R}}(p^2)\right]
 \left[1+\Sigma^{\bar\Pe\Pe}_{\mathrm{L}}(p^2)\right]},
\eeq
and $\Sigma'(p^2) \equiv \partial\Sigma(p^2)/\partial p^2$.
Here we have ignored terms with explicit factors $p^\mu$ or $p^\nu$,
because they turn out to be irrelevant for the $S$-matrix element, which
is to be constructed in the last step. 

The particles of the Compton process are labelled according to
\beq
\Pem(p,\kappa) \;\;+\;\; \gamma(k,\lambda) \;\;\longrightarrow\;\;
\Pem(p',\kappa') \;\;+\;\; \gamma(k',\lambda'),
\label{eq:comp}
\eeq
where the parentheses contain the momenta $p$, $k$, $p'$, $k'$
and the polarizations $\kappa$, $\lambda$, $\kappa'$, $\lambda'$ of the
respective particles. In the low-energy limit we have $p'=p$
and $k=k'=0$.
We denote the 
electron spinors by $u(p,\kappa)$ and $\bar u'(p,\kappa')$, and the
polarization vectors of the photons by $\veps(\lambda)$ and
$\veps^{\prime *}(\lambda)$. 
The $S$-matrix element is directly obtained from \refeq{eq:GAAee3} by
going on shell with the electrons and multiplying with the 
corresponding wave functions. 
At this point, the on-shell renormalization \cite{bgflong}
(see also \citere{onren})  comes into play. Within the on-shell scheme,
$\Me^2$ is identified with the pole position in the 
electron propagator, and the wave-function renormalization of the right- 
and left-handed field components fixes the residue of the pole to 1.
In this context, it should be noted that the electron self-energies 
(and their derivatives)
at $p^2=\Me^2$ are real quantities, since the electron is a stable particle. 
This implies that the pole of the propagator lies on the real axis.
In terms of the renormalized self-energies
$\Sigma^{\bar\Pe\Pe}_{\mathrm{R,L,S}}$, the three on-shell conditions imply 
\beq
 \Sigma^{\bar\Pe\Pe}_{\mathrm{S}}(\Me^2) =
-\Sigma^{\bar\Pe\Pe}_{\mathrm{R}}(\Me^2) =
-\Sigma^{\bar\Pe\Pe}_{\mathrm{L}}(\Me^2) =
\Me^2\left[
  \Sigma^{\prime\bar\Pe\Pe}_{\mathrm{R}}(\Me^2)
 +\Sigma^{\prime\bar\Pe\Pe}_{\mathrm{L}}(\Me^2)
+2\Sigma^{\prime\bar\Pe\Pe}_{\mathrm{S}}(\Me^2)
\right].
\label{eq:Sions}
\eeq
We recall that the photon wave function is already properly
renormalized, since the on-shell charge renormalization 
\cite{bgflett,bgflong,li95,bgfet}
fixes the residue of the photon propagator to 1.
Making use of the relations \refeq{eq:Sions} 
and the Dirac equations for the electron
spinors, all higher-order corrections in 
$G^{\mu\nu}_{\Ahat\Ahat\bar\Pe\Pe}$ of \refeq{eq:GAAee3} drop out in
the on-shell limit. The terms in \refeq{eq:GAAee3} which contain
explicit factors $p^\mu$ or $p^\nu$ vanish after contraction with the
polarization vectors $\veps$ and $\veps^{\prime *}$, because we can
choose their gauge such that $\veps\cdot p=\veps^{\prime *}\cdot p=0$.
The final result is
\beqar
\langle \Pem(p,\kappa'), \gamma(0,\lambda') |\,S\,|
\Pem(p,\kappa), \gamma(0,\lambda) \rangle 
&=& 
\veps^{\prime *}_\nu(\lambda') \,
\bar u(p,\kappa') G^{\mu\nu}_{\Ahat\Ahat\bar\Pe\Pe}(0,0,-p,p) u(p,\kappa) 
\, \veps_\mu(\lambda) 
\nn\\
&=& \frac{\ri e^2}{\Me}\,\bar u(p,\kappa')u(p,\kappa)\,
\veps(\lambda)\cdot\veps^{\prime *}(\lambda'),
\label{eq:eaea}
\eeqar
which is the usual Thomson scattering amplitude. Averaging over the
electron polarization, which is conserved, this yields the well-known
Thomson cross-section
\beq
\biggl(\frac{\rd\sigma}{\rd\Omega}\biggr)_{\mathrm{Thomson}} 
= \frac{\alpha^2}{\Me^2}\,
\left| \veps(\lambda)\cdot\veps^{\prime *}(\lambda') \right|^2,
\label{eq:thcs}
\eeq
where $\alpha=e^2/4\pi$ denotes the fine-structure constant.

Before turning to the generalization of the above reasoning, it is
mandatory to consider the question of photonic bremsstrahlung and the IR
problem. 
We have implicitly assumed that IR divergences are regulated in a
gauge-invariant manner, i.e.\ that the IR regulation preserves the Ward
identities. Usually an infinitesimally small photon mass $m_\gamma$
is introduced in actual calculations, which preserves the Ward
identities up to terms of ${\cal O}(m_\gamma)$.
The above derivation has shown that all virtual corrections vanish in
the low-energy limit before the IR regulation is released. Indeed this
is not the usual order in which kinematical limits of physical
observables are taken. Taking these two limits in the same order for the 
bremsstrahlung, one observes that photon emission is impossible in 
the low-energy limit, because there is no phase space. In other words, the
bremsstrahlung correction to Thomson scattering vanishes.
This conclusion is based on the assumption that the exchange of the
order of the two limits is allowed. 
This step can be justified by inspecting the form of the 
IR divergences at finite energies, which is explicitly known from QED 
\cite{ye61}, where the situation is the same for the Compton process.
Indeed it turns out that the logarithmically IR-divergent terms can be
separated order by order in the low-energy expansion. 
For a more detailed discussion of this subject, see \citere{bj65}. 

The one-loop corrections and the corresponding photonic
bremsstrahlung to Compton scattering are known both for QED \cite{br52} 
and for the electroweak SM \cite{rcegeg}.
In agreement with the considered theorem
these corrections vanish in the low-energy limit, which can be
explicitly checked by inspecting the analytical results given in the
literature.

\section{Generalization to other charged particles}
\label{se:other}

The generalization of the above derivation to the scattering between a photon
and any fermion $f$ with charge $Q_f e$ is quite obvious. One merely has to
replace the electron field $\Pe$ by $f$ and the charge $e$ by $-Q_f e$
in each step. While the above treatment remains exactly valid up to 
\refeq{eq:GAAee3} for any fermion $f$, differences occur when constructing 
the $S$-matrix element from the amputated Green function if $f$ is unstable. 
In this case the $f$ propagator possesses a complex pole, the position of 
which is not identical with the renormalized mass parameter $m_f^2$ in the 
on-shell renormalization scheme \cite{bgflong,onren}. 
It is well-known that this difference starts
to play a role at the two-loop level so that {\it mutatis mutandis} 
\refeq{eq:eaea} 
is valid for all fermions $f$ at least in the one-loop order. 
In this approximation also the above remarks on the photonic bremsstrahlung
corrections and the IR problem apply.
The generalization of the theorem to higher orders requires a proper
definition of on-shell renormalization and of the $S$-matrix 
for unstable particles \cite{ve63}, which is a highly non-trivial subject 
and clearly out of the scope of this article.

Next we turn to the process $\PWp\gamma\to\PWp\gamma$ in the SM,
where we meet analogous problems owing to the instability of the
W~bosons. Although the derivation of the low-energy theorem proceeds along 
the same lines as in the electron case, we again consider the single steps 
of the calculation in order to see where the restriction to the one-loop
level becomes relevant. Moreover,
the example of the W~boson nicely illustrates the elegance of the BFM.

In a general 't~Hooft gauge for the background
W-boson field, the relevant Ward identities are more involved owing to
the mixing between the background gauge fields $\What^\pm$ and the
background Goldstone fields $\phihat^\pm$. Therefore, we choose the
unitary gauge for the $\What^\pm$ fields, which is formally reached by
sending the background gauge parameter $\hat\xi_W$ to infinity. In this
limit the $\phihat^\pm$ fields completely drop out. We recall that
imposing the unitary gauge for a background gauge field in the BFM does
not lead to any difficulties with renormalizability, as only tree-level
quantities are concerned. This is in contrast to the conventional
quantization procedure, where in general Green functions are
non-renormalizable in the unitary gauge.

The desired Ward identity for the Green function
$G^{\Ahat\Ahat\What^+\What^-}_{\mu\nu\rho\sigma}$ is again obtained from
the first relation of Eq.~(11) in \citere{bgfet}. Note that in this
equation the unitary gauge for the $\What^\pm$ fields is reached by
disregarding all terms involving $\phihat^\pm$ and taking 
$\hat\xi_W\to\infty$. The final Ward identity possesses the same
structure as \refeq{eq:kGAAee2} for the electron case:
\beqar
\lefteqn{
\frac{\ri k_1^2}{\xihat_A}\, \frac{\ri k_2^2}{\xihat_A}\, k_1^\mu k_2^\nu 
G^{\Ahat\Ahat\What^+\What^-}_{\mu\nu\rho\sigma}(k_1,k_2,k_+,k_-) 
} \hspace{4em}
&& \nn\\ & = & 
 e^2 G^{\What^+\What^-}_{\rho\sigma}(k_+,-k_+)
-e^2 G^{\What^+\What^-}_{\rho\sigma}(k_+ +k_1,k_- +k_2)
\nn\\ && {}
-e^2 G^{\What^+\What^-}_{\rho\sigma}(k_+ +k_2,k_- +k_1)
+e^2 G^{\What^+\What^-}_{\rho\sigma}(-k_-,k_-).
\label{eq:kGAAWW1}
\eeqar
The amputation of the external photon propagators according to
\refeq{eq:amp} and \refeq{eq:GAXwi} leads to
\beq
\frac{\ri k_1^2}{\xihat_A}\, \frac{\ri k_2^2}{\xihat_A}\, k_1^\mu k_2^\nu 
G^{\Ahat\Ahat\What^+\What^-}_{\mu\nu\rho\sigma}(k_1,k_2,k_+,k_-) =
k_{1,\mu} k_{2,\nu}
G^{\mu\nu,\What^+\What^-}_{\Ahat\Ahat,\rho\sigma}(k_1,k_2,k_+,k_-).
\label{eq:kGAAWW2}
\eeq
The low-energy limit of
$G^{\mu\nu,\What^+\What^-}_{\Ahat\Ahat,\rho\sigma}$ is obtained by taking
derivatives from \refeq{eq:kGAAWW1} and \refeq{eq:kGAAWW2} with respect 
to $k_{1,\mu}$ and $k_{2,\nu}$,
\beq
G^{\mu\nu,\What^+\What^-}_{\Ahat\Ahat,\rho\sigma}(0,0,k_+,-k_+) =
e^2 \frac{\partial^2}{\partial k_{+,\mu}\partial k_{+,\nu}}
G^{\What^+\What^-}_{\rho\sigma}(k_+,-k_+).
\label{eq:GAAWW}
\eeq
In the unitary gauge for the background W-boson fields, the amputation of
each external $\What^\pm$ leg simply amounts to the multiplication of
$G^{\mu\nu,\What^+\What^-}_{\Ahat\Ahat,\rho\sigma}$ with the inverse of
the corresponding $\What^\pm$ propagator
$G^{\What^+\What^-}_{\alpha\beta}$, which is
the negative of the $\What^\pm$ two-point function
$\Gamma^{\What^+\What^-}_{\alpha\beta}$,
\beq
\left[ G^{\What^+\What^-}_{\alpha\beta}(k_+,-k_+) \right]^{-1} =
-\Gamma^{\What^+\What^-}_{\alpha\beta}(k_+,-k_+).
\eeq
Hence we have
\beqar
\lefteqn{
G^{\mu\nu\rho\sigma}_{\Ahat\Ahat\What^+\What^-}(0,0,k_+,-k_+) }
\nn\\ && =
e^2 \Gamma^{\What^+\What^-,\rho\alpha}(k_+,-k_+)
\left[ \frac{\partial^2}{\partial k_{+,\mu}\partial k_{+,\nu}}
G^{\What^+\What^-}_{\alpha\beta}(k_+,-k_+) \right]
\Gamma^{\What^+\What^-,\beta\sigma}(k_+,-k_+).
\label{eq:GAAWW2}
\eeqar
The higher-order corrections to $\Gamma^{\What^+\What^-}_{\alpha\beta}$
are represented by the transverse and longitudinal self-energies
$\Sigma^{\What^+\What^-}_\rT$ and $\Sigma^{\What^+\What^-}_\rL$,
respectively,
\beqar
\Gamma^{\What^+\What^-}_{\alpha\beta}(k_+,-k_+) &=& 
-\ri\biggl( g_{\alpha\beta}-\frac{k_{+,\alpha}k_{+,\beta}}{k_+^2} \biggr)
\left[ k_+^2-\MW^2+\Sigma^{\What^+\What^-}_\rT(k_+^2) \right]
\nn\\ && {}
+\ri\frac{k_{+,\alpha}k_{+,\beta}}{k_+^2}
\left[ \MW^2-\Sigma^{\What^+\What^-}_\rL(k_+^2) \right].
\eeqar
In terms of these self-energies, \refeq{eq:GAAWW2} reads
\beqar
\lefteqn{
G^{\mu\nu\rho\sigma}_{\Ahat\Ahat\What^+\What^-}(0,0,k_+,-k_+) }
\nn\\ & = & 
-2\ri e^2 g^{\mu\nu}g^{\rho\sigma}
\left[ 1+\Sigma^{\prime\What^+\What^-}_\rT(k_+^2) \right]
\nn\\ && {}
-\frac{\ri e^2}{k_+^2} (g^{\mu\rho}g^{\nu\sigma}+g^{\mu\sigma}g^{\nu\rho})
\left[ k_+^2-\MW^2+\Sigma^{\What^+\What^-}_\rT(k_+^2) \right]
\frac{k_+^2+\Sigma^{\What^+\What^-}_\rT(k_+^2)
-\Sigma^{\What^+\What^-}_\rL(k_+^2)}
{\MW^2-\Sigma^{\What^+\What^-}_\rL(k_+^2)}
\nn\\ && {} + \;\; 
\left( \mbox{terms involving $k_+^\mu$, $k_+^\nu$, $k_+^\rho$, or
$k_+^\sigma$ } \right).
\label{eq:GAAWW3}
\eeqar
Up to this point no restriction to the one-loop level is necessary.

The $S$-matrix element of the process
\beq
\PWp(k_+,\lambda_+) \;\;+\;\; \gamma(k,\lambda) \;\;\longrightarrow\;\;
\PWp(k'_+,\lambda'_+) \;\;+\;\; \gamma(k',\lambda')
\label{eq:awaw}
\eeq
for low-energetic photons, i.e.\ $k=k'=0$ and $k'_+=k_+$, is obtained
from $G^{\mu\nu\rho\sigma}_{\Ahat\Ahat\What^+\What^-}$ in \refeq{eq:GAAWW3}
by taking $k_+^2\to\MW^2$, contraction with the respective
polarization vectors $\veps_{+,\rho}(\lambda_+)$, $\veps_\mu(\lambda)$, 
$\veps^{\prime *}_{+,\sigma}(\lambda'_+)$, $\veps^{\prime *}_\nu(\lambda')$,
and multiplication with the wave-function correction factor 
$\sqrt{R_\What}$ for each external $\What^\pm$ field \cite{bgfet},
\beqar
\lefteqn{
\langle \PWp(k_+,\lambda'_+), \gamma(0,\lambda') |\,S\,|
\PWp(k_+,\lambda_+), \gamma(0,\lambda) \rangle } \hspace{4em}
\nn\\[.3em] && =
R_\What \, \veps^{\prime *}_{+,\sigma}(\lambda'_+) \, 
\veps^{\prime *}_\nu(\lambda') \,
G^{\mu\nu\rho\sigma}_{\Ahat\Ahat\What^+\What^-}(0,0,k_+,-k_+) \,
\veps_{+,\rho}(\lambda_+) \, \veps_\mu(\lambda).
\label{eq:awaw2}
\eeqar
The parts of $G^{\mu\nu\rho\sigma}_{\Ahat\Ahat\What^+\What^-}$
containing explicit $k_+^\mu$ factors etc., which are already suppressed
in \refeq{eq:GAAWW3}, do not contribute to the $S$-matrix element because 
of the transversality of the W-boson polarization vectors, 
$\veps_+\cdot k_+ = \veps_+^{\prime *}\cdot k_+ = 0$, and the gauge
choice for the photon polarization vectors, 
$\veps\cdot k_+ = \veps^{\prime *}\cdot k_+ = 0$. 
For the other terms in \refeq{eq:GAAWW3} we have to inspect the explicit
form of the renormalization conditions. 
In the BFM renormalization scheme of \citere{bgflong}, the 
W-boson mass is determined by the usual on-shell condition \cite{onren}
\beq
\Re\left\{\Sigma^{\What^+\What^-}_\rT(\MW^2)\right\} = 0
\eeq
for the transverse part of the renormalized self-energy.
The wave-function correction factor $R_\What$ \cite{bgfet} is given by
\beq
R_\What = \left[ 
1+\Re\left\{\Sigma^{\prime\What^+\What^-}_\rT(\MW^2)\right\} \right]^{-1}.
\eeq
Making use of these relations when going on shell in \refeq{eq:GAAWW3},
one finds 
\beqar
\lefteqn{
\langle \PWp(k_+,\lambda'_+), \gamma(0,\lambda') |\,S\,|
\PWp(k_+,\lambda_+), \gamma(0,\lambda) \rangle } \hspace{4em}
\nn\\[.3em] && =
-2\ri e^2 
\veps^{\prime *}_{+,\sigma}(\lambda'_+) \, 
\veps^{\prime *}_\nu(\lambda') \,
\veps_{+,\rho}(\lambda_+) \, \veps_\mu(\lambda) \,
\left[ g^{\mu\nu}g^{\rho\sigma} 
+ \ri\Re\left\{{\cal O}(e^2)\right\} \right],
\label{eq:awaw3}
\eeqar
which implies that the corrections of relative order ${\cal O}(e^2)$
only yield an unobservable phase factor, because they do not interfere
with the Born amplitude. Therefore, the squared $S$-matrix element reads
\beqar
\lefteqn{
\left| \langle \PWp(k_+,\lambda'_+), \gamma(0,\lambda') |\,S\,|
\PWp(k_+,\lambda_+), \gamma(0,\lambda) \rangle \right|^2 
} \hspace{4em}
\nn\\[.3em] && =
4e^4\, \big|\veps_+(\lambda_+)\cdot\veps_+^{\prime *}(\lambda'_+)\big|^2 \, 
\big|\veps(\lambda)\cdot\veps^{\prime *}(\lambda')\big|^2
\;+\; {\cal O}(e^8),
\eeqar
leading to the Thomson cross-section \refeq{eq:thcs} after averaging
over the (conserved) W-boson polarization.
The bremsstrahlung corrections of relative order ${\cal O}(e^2)$ also
vanish, as in the cases considered above. This is due to the fact that
the IR behaviour of these corrections is exactly the same as for
fermions (see for instance the last paper of \citere{onren}).

Finally, we point out that it is straightforward to carry over the above 
reasoning, which was explained in detail for the reactions 
$\Pem\gamma\to\Pem\gamma$ and $\PWp\gamma\to\PWp\gamma$, to other charged 
particles in various extensions of the SM. For instance, adding a scalar
SU(2)$\times$U(1) multiplet containing the charged boson $\PS^\pm$ with
electric charge $\pm Q_\PS$, one obtains
\beqar
\left| \langle \PS^\pm(k), \gamma(0,\lambda') |\,S\,|
\PS^\pm(k), \gamma(0,\lambda) \rangle \right|^2 =
4Q_{\PS}^4e^4\, \big|\veps(\lambda)\cdot\veps^{\prime *}(\lambda')\big|^2
\;+\; {\cal O}(e^8).
\eeqar
for $\PS^\pm\gamma\to\PS^\pm\gamma$ at low energies.

\section{Summary}
\label{se:sum}

We have proved that the electroweak radiative corrections to Compton
scattering vanish in the low-energy limit in all orders of perturbation
theory. This generalizes Thirring's
theorem for QED to the electroweak Standard Model. The theorem
guarantees that the on-shell definition of the electric charge of the
electron corresponds to the electromagnetic coupling measured via
Thomson scattering. An analogous theorem for photon scattering off
W~bosons has also been derived. However, the instability of the W~bosons
restricted the simple formulation of the theorem to the
one-loop level.

Since these low-energy theorems are a consequence of gauge invariance
and on-shell renormalization, their generalization from Abelian gauge
theories such as QED to non-Abelian ones is non-trivial. We have formulated
our derivation in the framework of the background-field method, rendering
the formal proof very simple and transparent, because 
vertex and Green functions obey, in this formalism, QED-like Ward identities.

\section*{Acknowledgement}

The author thanks A.~Denner and H.~Spiesberger for useful discussions
and a careful reading of the manuscript.

\def\vol#1{{\bf #1}}
\def\mag#1{{\sl #1}}

\end{document}